\documentclass[fleqn,10pt]{wlscirep}
\title{Macroscopic phase separation of superconductivity and ferromagnetism in Sr$_{0.5}$Ce$_{0.5}$FBiS$_{2-x}$Se$_x$ revealed by $\mu$SR}

\author[1,$\dagger$]{A.M. Nikitin}
\author[2,3]{V. Grinenko}
\author[2]{R. Sarkar}
\author[4]{J.-C. Orain}
\author[1]{M.V. Salis}
\author[1]{J. Henke}
\author[1]{Y.K. Huang}
\author[2]{H.-H. Klauss}
\author[4]{A. Amato}
\author[1,*]{A. de Visser}

\affil[1]{Van der Waals - Zeeman Institute, University of Amsterdam, 1098 XH Amsterdam, The Netherlands}
\affil[2]{Institute of Solid State and Materials Physics, Technical University Dresden, 01062 Dresden, Germany}
\affil[3]{Leibniz Institute for Solid State and Materials Research (IFW), 01069 Dresden, Germany}
\affil[4]{Laboratory for Muon-Spin Spectroscopy, Paul Scherrer Institute, 5232 Villigen PSI,  Switzerland}
\affil[$\dagger$]{current address: Laboratory for Muon-Spin Spectroscopy, PSI,  Switzerland; artem.nikitin@psi.ch}
\affil[*]{a.devisser@uva.nl}

\begin{abstract}
The compound Sr$_{0.5}$Ce$_{0.5}$FBiS$_{2}$ belongs to the intensively studied family of layered BiS$_2$ superconductors. It attracts special attention because superconductivity at $T_{sc} = 2.8$~K was found to coexist with local-moment ferromagnetic order with a Curie temperature $T_C = 7.5$~K. Recently it was reported that upon replacing S by Se $T_C$ drops and ferromagnetism becomes of an itinerant nature (Thakur \textit{et al}., Sci. Reports 6, 37527 (2016)). At the same time $T_{sc}$ increases and it was argued superconductivity coexists with itinerant ferromagnetism. Here we report a muon spin rotation and relaxation study ($\mu$SR) conducted to investigate the coexistence of superconductivity and ferromagnetic order in Sr$_{0.5}$Ce$_{0.5}$FBiS$_{2-x}$Se$_x$ with $x=0.5$ and $1.0$. By inspecting the muon asymmetry function we find that both phases do not coexist on the microscopic scale, but occupy different sample volumes. For $x=0.5$ and $x=1.0$ we find a ferromagnetic volume fraction of $\sim \, 8~\%$ and $\sim \, 30~\%$ at $T=0.25$~K, well below $T_{C} = 3.4$~K and $T_C = 3.3$~K, respectively. For $x=1.0$ ($T_{sc} = 2.9$~K) the superconducting phase occupies the remaining sample volume ($\sim \, 70~\%$), as shown by transverse field experiments that probe the Gaussian damping due to the vortex lattice. We conclude ferromagnetism and superconductivity are macroscopically phase separated.
\end{abstract}

\begin{document}

\flushbottom
\maketitle
\thispagestyle{empty}

\noindent

\section*{Introduction}

The interplay between superconductivity and magnetism has been a central issue in superconductivity research for several decades now. Especially, the idea that superconductivity and ferromagnetism can occur simultaneously has attracted the attention of researchers throughout the years. Already in 1957 Ginzburg argued a superconducting phase can exist in a ferromagnet when the spontaneous magnetization $M_0$ is smaller than the lower critical field $\mu _0 H_{c1}$, but also pointed out the "almost complete impossibility in practice, under ordinary conditions, to observe superconductivity in any sort of ferromagnets"~\cite{Ginzburg1957}. Two years later Anderson and Suhl asserted that a ferromagnetic alignment of spins in a superconductor can occur, but only in a very small domain-like `cryptoferromagnetic' configuration, where the domain size $l_D$ is smaller than the superconducting coherence length $\xi$~\cite{Anderson&Suhl1959}. On the other hand, early experimental work on Gd doped La~\cite{Matthias1958a} and (Ce,Gd)Ru$_2$ alloys~\cite{Matthias1958b} indicated superconductivity and ferromagnetism are competing phenomena, which was subsequently corroborated by studies of Chevrel phases, such as ErRh$_4$B$_4$, where superconductivity is expelled when ferromagnetic order sets in~\cite{Fertig1977}. Here the general idea is that the ferromagnetic exchange field impedes the formation of spin-singlet Cooper pairs that is prescribed by the microscopy theory of Bardeen, Cooper and Schrieffer (BCS)~\cite{Berk&Schrieffer1966}. Notwithstanding this restriction, the search for ferromagnetic superconductors continued unremittingly. This resulted in the discovery of perhaps a dozen of remarkable materials in which superconductivity and ferromagnetism exhibit coexistence. However, in most of these systems, superconductivity and ferromagnetism are confined to different crystallographic planes (e.g. RuSr$_2$GdCu$_2$O$_8$~\cite{Bernhard1999}) and/or to different electron subsystems, i.e. conduction \textit{and} local $4f$ magnetic moments (e.g. ErNi$_2$B$_2$C~\cite{Canfield1996}, EuFe$_2$(As,P)$_2$~\cite{Liu2016} and RbEuFe$_4$As$_4$~\cite{Cao2011}). Also, some of the systems have metallurgical difficulties (Y$_9$Co$_7$~\cite{Bochenek2015,Larson_thesis}) or exhibit a form of phase separation (e.g. the electron gas at the SrTiO$_3$/LaAlO$_3$ interface~\cite{Bert2011}). On the contrary, in a small group of uranium-based correlated metals formed by UGe$_2$ (under pressure~\cite{Saxena2000}), URhGe~\cite{Aoki2001} and UCoGe~\cite{Huy2007}, ferromagnetism and superconductivity do coexist on the microscopic scale and are carried by the same $5f$ electrons. This is corroborated by the itinerant nature of the ferromagnetic state. The superconducting transition temperature, $T_{sc}$, is below the Curie temperature $T_C$, hence the label superconducting ferromagnets. Superconducting ferromagnets have provided new opportunities to investigate exotic superconductivity. Theoretical work predicts an odd-parity Cooper pair state mediated by longitudinal spin fluctuations~\cite{Fay&Appel1980,Monthoux2007}.

In a recent publication, Thakur \textit{et al.}~\cite{Thakur2016} provide evidence that Sr$_{0.5}$Ce$_{0.5}$FBiS$_{2-x}$Se$_x$, with $x=0.5$ and $1.0$, is a new superconducting ferromagnet ($T_{sc} < T_C$). Magnetization measurements for $x \geq 0.5$ signal bulk superconductivity and a small average ordered Ce-moment ($\sim ~0.1~\mu_B$) in the superconducting state, which is in line with itinerant ferromagnetism. This is further substantiated by specific heat measurements for $x=0.5$ that show the magnetic entropy, $S_m$, per Ce atom is only 4~\% of the expected value for trivalent Ce-$4f$ ($J=5/2$), $S_m = 0.04 \times$Rln6. Moreover, they report a dual and quite unusual hysteresis loop in the magnetization below $T_{sc}$ corresponding to the coexistence of ferromagnetism and superconductivity. The parent material Sr$_{0.5}$Ce$_{0.5}$FBiS$_2$ has a ferromagnetic transition at $T_C = 7.5$~K and superconducts at $T_{sc} = 2.8$~K~\cite{Lin2015}. Here magnetic order is due to local moments, as evidenced by their magnitude ($\sim ~1~\mu_B$) and the large value of the magnetic entropy $S_m = 0.5 \times$Rln2, assuming a doublet crystal field ground state ($J=1/2$). In their recent report on the Sr$_{0.5}$Ce$_{0.5}$FBiS$_{2-x}$Se$_x$ system Thakur \textit{et al.} establish that upon replacing S by isovalent Se $T_C$ reduces and $T_{sc}$ is enhanced. At the same time magnetization and specific heat  data were interpreted as evidence for the itinerant character of the $4f$-electrons at high Se doping. Consequently they argue magnetic order and superconductivity are carried by the same type of electrons for $x \geq 0.5$. This and the small size of the  ordered moment leads them to draw a close parallel between Sr$_{0.5}$Ce$_{0.5}$FBiS$_{2-x}$Se$_x$ and UCoGe as regards the coexistence of superconductivity and itinerant ferromagnetism.

Here we report muon spin relaxation and rotation ($\mu$SR) experiments on Sr$_{0.5}$Ce$_{0.5}
$FBiS$_{2-x}$Se$_x$ conducted to investigate the coexistence of superconductivity and ferromagnetism on the microscopic scale. $\mu$SR is the technique \textit{par excellence} to probe small magnetic moments, as well as to determine the superconducting and magnetic volume fractions in a crystal~\cite{Amato1997}. The latter information can normally not be extracted from macroscopic measurements such as the magnetization or specific heat. The experiments were performed on two samples with different Se content $x$: (\textit{i}) $x=0.5$, this sample is taken from the same batch as used by Thakur \textit{et al.}~\cite{Thakur2016}, and (\textit{ii}) $x=1.0$, this sample was synthesized  at the University of Amsterdam. We have carried out zero field and transverse field (TF = 10~mT) $\mu$SR experiments in the temperature range 0.25-10 K. We detect both the ferromagnetic order and superconductivity. However, by inspecting the muon asymmetry function we conclude these ordered states do not coexist on the microscopic scale, but occupy different sample volumes. For $x=0.5$ and $x=1.0$ we find the ferromagnetic volume fraction is $\sim \, 8~\%$ and $\sim \, 30~\%$ at $T=0.25$~K, \textit{i.e.} well below $T_C = 3.4$~K and $T_C = 3.3$~K, respectively. Transverse field experiments carried out for $x=1.0$ demonstrate the superconducting phase ($T_{sc} = 2.9$~K) occupies the remaining sample volume ($\sim \, 70~\%$).

\section*{Results and Analysis}

\textbf{Sr$_{0.5}$Ce$_{0.5}$FBiSSe} A polycrystalline sample was prepared at the University of Amsterdam and characterized by dc magnetization, ac susceptibility, electrical resistivity and specific heat, as shown in Figures S1-S6 in the Supplementary Information (SI) file. The resulting $T_C = 3.3$~K and $T_{sc} = 2.9$~K are in excellent agreement with the values reported by Thakur \textit{et al.} for the same Se content ($x=1.0$). At $T = 2.0$~K the low-field magnetization data point to a small ordered moment of $\sim \, 0.2~\mu_B /$Ce. By increasing the field the moment grows and saturates in the high field region (9~T) at a large value of $0.9~\mu_B /$Ce. Ac-susceptibility measurements shows a large diamagnetic signal, which implies a superconducting screening fraction of $\sim \, 0.7$. The superconducting state is further characterized by the electrical resistivity in an applied magnetic field (Fig. S5) and dc-magnetization measurements (Fig. S6). The data were used to extract a lower critical field, $B_{c1} = 0.6$~mT (Fig.~S6c), and an upper critical field $B_{c2}=2.9$~T for $T \rightarrow 0$ (Fig.~S5). All in all, these results show our sample has very similar magnetic and superconducting properties as the sample with $x=1.0$ investigated by Thakur \textit{et al.}~\cite{Thakur2016}. In the following two sections we present the results of the $\mu$SR experiments for $x=1.0$.

\subsubsection*{Zero field experiments}

\begin{figure}[ht]
\centering
\includegraphics[width=\linewidth]{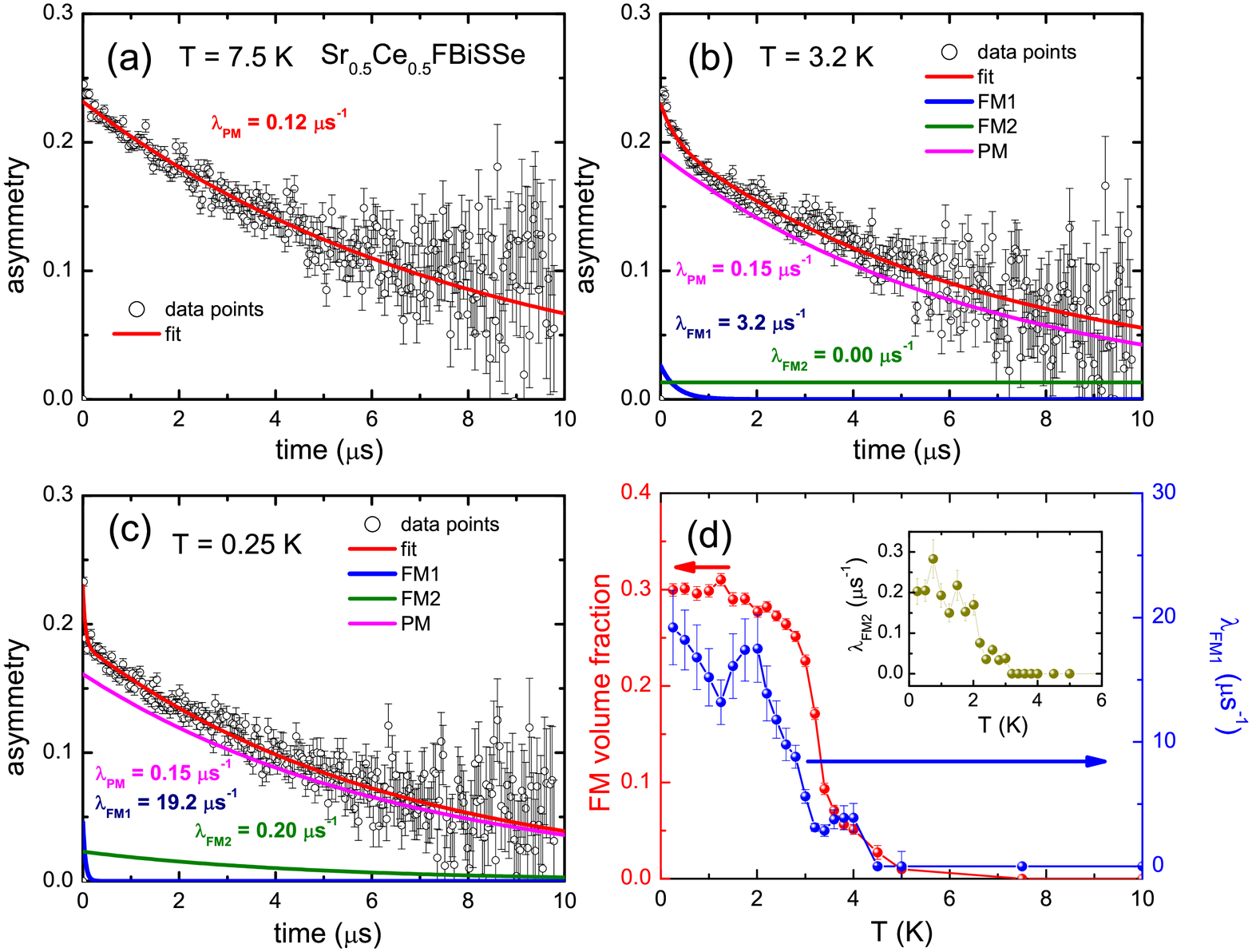}
\caption{Zero field $\mu$SR data measured for Sr$_{0.5}$Ce$_{0.5}$FBiSSe. Panels (a), (b) and (c): Asymmetry as a function of time at temperatures of 7.5~K, 3.2~K and 0.25~K, respectively. The red lines are fits to the muon depolarization function eq.~1. The blue, magenta and green lines are the contributions from the ferromagnetic fast (FM$_1$), ferromagnetic slow (FM$_2$), and paramagnetic (PM) signals, respectively. The corresponding relaxation rates are listed. Panel (d): Temperature variation of the ferromagnetic volume fraction $f_{FM}$ (red symbols, left axis) and $\lambda_{FM_1}$ (blue symbols, right axis). Inset: Temperature variation of $\lambda_{FM_2}$.}
\label{figure:ZF}
\end{figure}

The muon ($\mu ^+$) depolarization in zero field was measured in the temperature range 0.25 - 10 K. Typical spectra in the time domain are shown in Fig.~\ref{figure:ZF}. In the paramagnetic state at $T=7.5$~K we observe a pronounced $\mu ^+$ depolarization indicating the presence of slow magnetic fluctuations. The signal has the full experimental asymmetry ($A_{tot}=0.24$) and accounts for the whole sample volume. Upon cooling to 3.2 K, \textit{i.e.} to just below $T_C$, an additional rapid depolarization component appears at short times, which we associate with the ferromagnetic phase. This component further develops with decreasing temperature and the corresponding relaxation rate increases and reaches a value of $\sim \, 18~\mu s ^{-1}$ at the lowest temperatures. The asymmetry associated with the ferromagnetic phase  (Fig.~\ref{figure:ZF}c) tells us the it occupies about 30~\% of the sample volume. This can be put on a firm footing by the analysis of the zero field data with the two-component $\mu ^+$ depolarization function
\begin{equation}
G(t) = A_{tot}[f_{FM} ( \frac{2}{3} e^{-\lambda_{FM_1} t} + \frac{1}{3} e^{-\lambda_{FM_2} t}) + f_{PM} e^{-\lambda_{PM} t}].
\end{equation}

Here $f_{FM}$ and $f_{PM}$ are the ferromagnetic (FM) and paramagnetic (PM) volume fractions, respectively, and $f_{FM}+f_{PM}=1$. $\lambda_{PM}$ is the relaxation rate in the paramagnetic phase, and $\lambda_{FM_1}$ and $\lambda_{FM_2}$ are the fast (2/3 component) and slow (1/3 component) relaxation rates in the ferromagnetic phase, respectively. When fitting the ferromagnetic contribution we fixed $\lambda _{PM}$ at $0.15~\mu s ^{-1}$ and fixed the total asymmetry $A_{tot} = 0.24$. We remark this value of $\lambda _{PM}$ is slightly larger than the value extracted at 7.5 K (see Fig.~\ref{figure:ZF}a), but it improved the quality of the fit. The results of this fitting procedure at 3 typical temperatures are shown in Figs.~\ref{figure:ZF}a,b,c. In Fig.~\ref{figure:ZF}d we show the temperature variation of $f_{FM}$ and of the relaxation rates $\lambda_{FM_1}$ and $\lambda_{FM_2}$. Clearly, $f_{FM}$ shows the strongest increase at $T_C =3.2$~K and then levels off to a ferromagnetic volume fraction of 30~\%. Correspondingly,  $\lambda_{FM_1}$ and $\lambda_{FM_2}$ increase and saturate in the ferromagnetic phase. We remark the ratio of the fast and slow relaxation rates is large, $\lambda_{FM_1}$/$\lambda_{FM_2} \approx 100$.

\subsubsection*{Transverse field experiments}

\begin{figure}[ht]
\centering
\includegraphics[width=\linewidth]{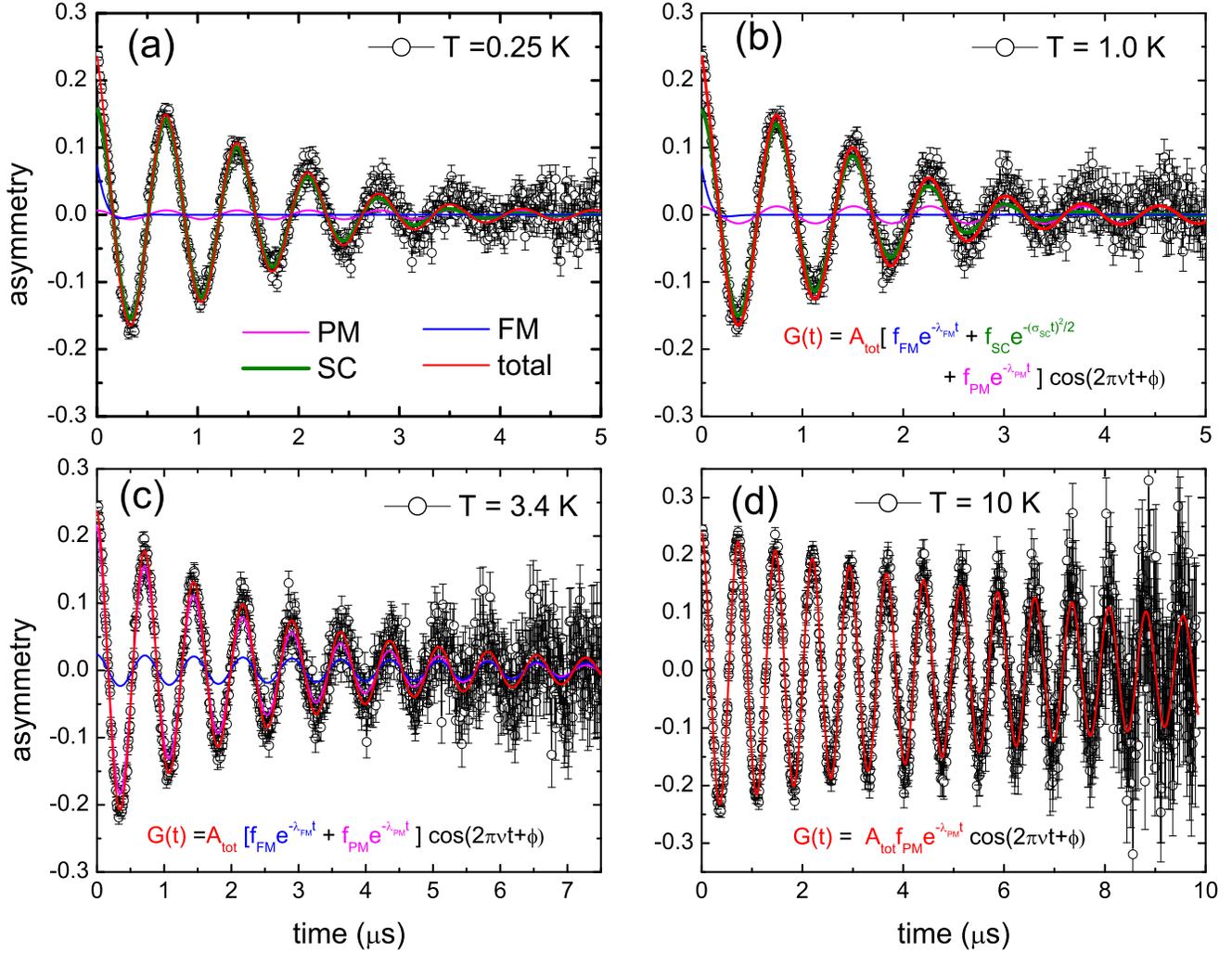}
\caption{Transverse field $\mu$SR data measured for Sr$_{0.5}$Ce$_{0.5}$FBiSSe. The applied field is $B$ = 10~mT. Panels (a), (b), (c) and (d): Asymmetry as a function of time at temperatures of 0.25~K, 1.0~K, 3.4~K and 10~K. The red lines are fits to the muon depolarization function eq.~2. The blue, magenta and green lines are the contributions from the ferromagnetic (FM), paramagnetic (PM) and superconducting (SC) signals, respectively. The muon depolarization functions of the contributing components are listed. }
\label{figure:TF}
\end{figure}

\begin{figure} [ht]
\includegraphics[width=\linewidth]{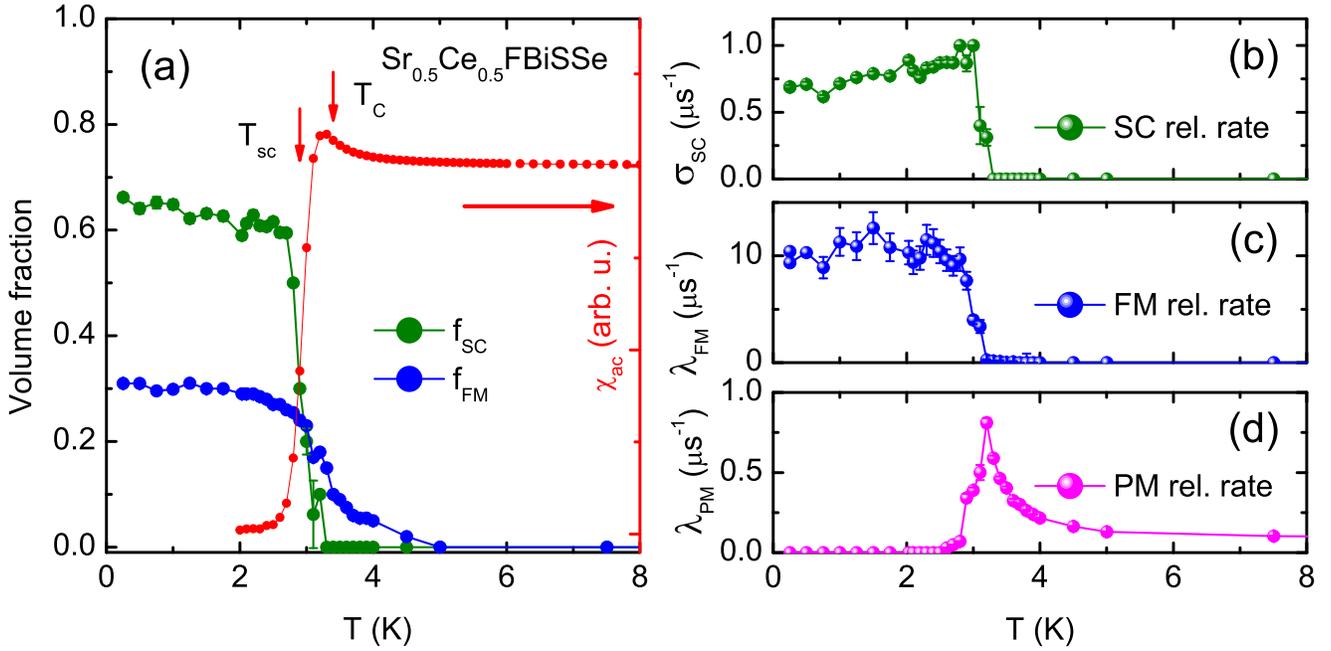}
\caption{Fit parameters of the analysis of the transverse field $\mu$SR data measured for Sr$_{0.5}$Ce$_{0.5}$FBiSSe. Panel (a): Temperature variation of the ferromagnetic $f_{FM}$ (blue symbols) and superconducting $f_{SC}$ (green symbols) volume fractions (left axis), and the ac-susceptibility (red symbols, right axis). $T_{sc}$ and $T_C$ extracted from $\chi _{ac}$ are indicated by arrows. Panel (b), (c) and (d): Temperature variation of the Gaussian damping rate due to superconductivity, and the exponential relaxation rates of the ferromagnetic and paramagnetic phases, respectively.}
\label{figure:fitparTF}
\end{figure}

Transverse field $\mu$SR measurements were carried out in a small magnetic field of 10~mT in the temperature range 0.25-10~K. Typical spectra are shown in Fig.~\ref{figure:TF}. In the paramagnetic phase a sizeable damping is observed with an exponential relaxation rate $\lambda_{PM} = 0.10~\mu s ^{-1}$ at 10~K. This value compares well to the value found in the zero field experiments. Upon approaching the Curie point the damping increases considerably, as shown in the spectrum at $T=3.4$~K (Fig.~\ref{figure:TF}c), while by further cooling to below $T_{sc}$ additional damping due to the flux line lattice appears (Figs.~\ref{figure:TF}a,b). Good fits to the transverse field $\mu$SR spectra are obtained with the three-component depolarization function

\begin{equation}
G(t) = A_{tot}[f_{SC}e^{\frac{-(\sigma_{SC} t)^2}{2}} + f_{FM} e^{-\lambda_{FM} t} + f_{PM} e^{-\lambda_{PM} t}] \cos(2\pi \nu t + \phi),
\end{equation}

where $f_{SC}$ is the superconducting volume fraction and the Gaussian damping due to the vortex lattice is expressed by the relaxation rate $\sigma_{SC}$. The parameters $f_{FM}$, $f_{PM}, \lambda_{FM}$ and $\lambda_{PM}$ have the same meaning as in the zero-field case. The muon precession frequency is given by $\nu$ and its phase by $\phi$. When analyzing the spectra at the lowest temperatures we first used eq.~2 with the  first two terms only ($f_{PM}=0$). This irrevocably showed that close to 30 \% of the sample exhibits ferromagnetic order, as was deduced from the zero field experiment, while superconductivity occupies the remaining 70 \% of the sample volume. We remark a slightly better fit was obtained by allowing for an additional small paramagnetic volume fraction with relaxation rate $\lambda_{PM}=0$ which accounts for an almost negligible 3~\% of the sample volume, $f_{PM} = 0.03$ (see Fig.~\ref{figure:TF}a). Next, in order to follow the temperature variation of the fit parameters of the different components we used the following constraints: (i) $f_{FM} (T)$ is taken equal to the values obtained in zero field (Fig.~\ref{figure:ZF}d), (ii) $\lambda_{PM} = 0$ for $T < 2$~K, and (iii) $\sigma_{SC} = 0$ for $T > 3$~K. The resulting fit parameters are shown in Fig.~\ref{figure:fitparTF}. For $T \rightarrow 0$ $\lambda_{FM} \approx 10 ~\mu s^{-1}$ attains a large value like in the zero field experiment. The superconducting state is characterized by a Gaussian damping with $\sigma_{SC} = 0.70 ~\mu s^{-1}$ for $T \rightarrow 0$. Upon increasing the temperature $\lambda_{FM}$ shows a smooth temperature variation and drops to zero at $T_C$. On the other hand $\sigma_{SC}$ first increases with increasing temperature and then drops to zero at $T_{sc}$. We remark this non-BCS increase is an artefact of the fitting procedure close to $T_{sc}$. Since $T_{sc} \approx T_C$ it is difficult to disentangle the three components in the vicinity of the phase transitions. $f_{FM}(T)$ and $f_{SC}(T)$ are traced in Fig.~\ref{figure:fitparTF}a. For comparison we have plotted in the same figure the ac-susceptibility measured on a piece of the same $x=1.0$ batch (see also Fig.~S4). The smooth variation of $f_{FM}(T)$ and $f_{SC}(T)$ to zero is in good agreement with $T_C = 3.3$~K and $T_{sc} = 2.9$~K extracted from the magnetic measurements.

\section*{}

\textbf{Sr$_{0.5}$Ce$_{0.5}$FBiS$_{1.5}$Se$_{0.5}$} A polycrystalline sample with $x=0.5$ was taken from the same batch as used in Ref.~\citenum{Thakur2016}. The $x=0.5$ compound has been characterized extensively by resistivity, ac-susceptibility, magnetization and specific heat measurements~\cite{Thakur2016}. DC magnetization measurements in an applied field of 1~mT were used to determine the transition temperatures $T_C = 3.5$~K and $T_{sc} = 2.7$~K. A small spontaneous moment was found with magnitude $\sim \, 0.09~\mu_B /$Ce at $T = 2$~K. Furthermore, the analysis of the specific heat data pointed to a low value for the magnetic entropy $S_m = 0.04 \times$Rln2 associated with the ferromagnetic transition. The small value of the ordered moment and the reduced entropy were taken as evidence for the development of itinerant magnetism upon replacing S by Se.

\subsubsection*{Zero field experiments}
\begin{figure}[ht]
\centering
\includegraphics[width=\linewidth]{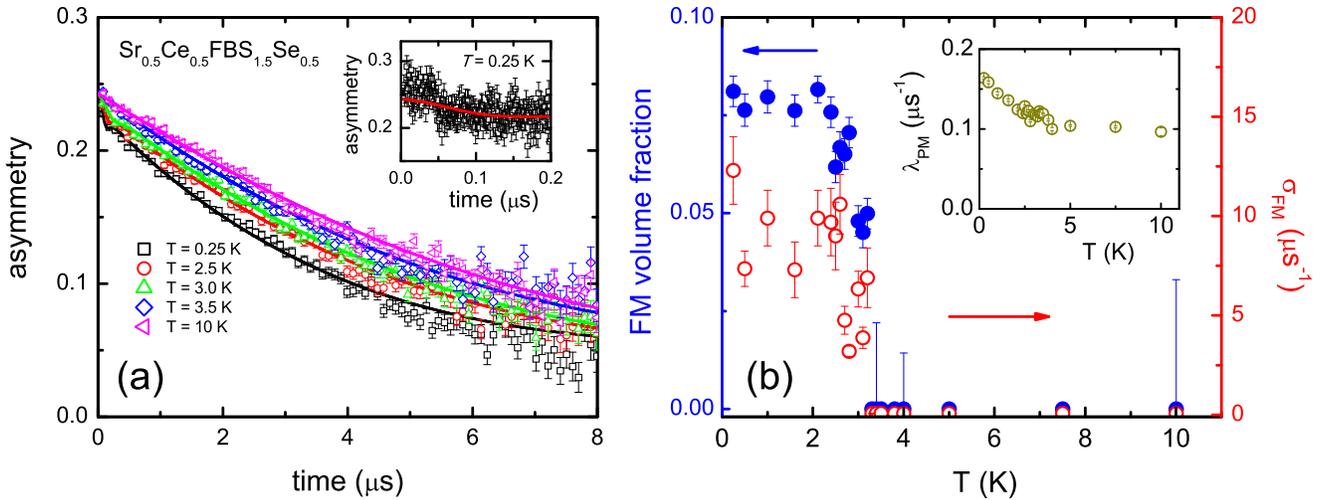}
\caption{Zero field $\mu$SR data for Sr$_{0.5}$Ce$_{0.5}$FBiS$_{1.5}$Se$_{0.5}$. Panel (a): Asymmetry as a function of time at temperatures as indicated. The solid lines are fits to eq.~3. The inset shows the asymmetry function at $T=0.25$~K up to $0.2~\mu$s$^{-1}$. Panel (b): Temperature variation of the ferromagnetic volume fraction $f_{FM}$ (blue symbols, left axis) and $\sigma_{FM}$ (red symbols, right axis). Inset: Temperature variation of $\lambda_{PM}$.}
\label{figure:ZF_d}
\end{figure}

Zero field $\mu$SR time spectra were taken in the temperature interval 0.25-10~K. The data at a few selected temperatures are shown in Fig.~\ref{figure:ZF_d}. In the paramagnetic state the muon depolarization is an exponential function of time with a similar relaxation rate as for $x=1.0$. At $T=0.25$~K, deep in the magnetic phase, an additional depolarization mechanism appears at small times ($t < 0.1 ~\mu$s$^{-1}$) (see also the inset in Fig.~\ref{figure:ZF_d}a), but it is not as pronounced as for $x=1.0$. An elaborate analysis showed it is due to a magnetic volume fraction of $\sim \, 0.08$ only. Best fits were obtained by using a two component depolarization function with: (i) fast relaxation due to a (disordered) ferromagnetic phase and (ii) exponential relaxation in the non-ferromagnetic part due to dilute magnetic impurities~\cite{Yaounc&Dalmas2011}:

\begin{equation}
 G(t) = A_{tot}[f_{FM} ( \frac{1}{3} + \frac{2}{3}[1 - (\sigma_{FM} t)^2]e^{\frac{-(\sigma_{FM} t)^2}{2}} )
 + f_{PM}( \frac{1}{3} + \frac{2}{3}[1 - (\sigma_{N} t)^2 - \lambda_{PM}t]e^{-(\frac{(\sigma_{N} t)^2}{2} - \lambda_{PM}t)})]
\end{equation}

Here $f_{FM}$ and $f_{PM} = 1 - f_{FM}$ are the ferromagnetic (FM) and paramagnetic (PM) volume fractions respectively, $\sigma_{FM}$ is the ferromagnetic relaxation rate, $\lambda_{PM}$ is the paramagnetic relaxation rate, and $\sigma_{N}$ is the nuclear contribution which was fixed at $0.07~\mu$s$^{-1}$. The fit results are shown by the solid lines in Fig.~\ref{figure:ZF_d}a. The temperature variation of $f_{FM}$, $\sigma_{FM}$ and $\lambda_{PM}$ is reported in Fig.~\ref{figure:ZF_d}b. The analysis clearly shows the ferromagnetic phase is bound to a volume fraction of $\sim \, 0.08$ only.

\section*{Discussion and concluding remarks}
The $\mu$SR data irrevocably show that the magnetism associated with the ordering temperatures $T_C = 3.4$~K and $T_C = 3.3$~K for $x=0.5$ and $x=1.0$, respectively, develops in a part of the sample only. This tells us that substituting Se for S in  Sr$_{0.5}$Ce$_{0.5}$FBiS$_{2}$ results in electronic phase separation. Moreover, our $\mu$SR analysis with large magnetic relaxation times points to a considerable amount of disorder in the magnetic phase. We stress that our conclusions are robust and do not depend on details of the fitting procedure used. These results sharply contrast with $\mu$SR spectra measured for the superconducting itinerant ferromagnet UCoGe with $T_{sc}=0.5$~K and $T_C= 3.0$~K~\cite{deVisser2009}. In this case a spontaneous muon precession frequency of 2 MHz ($T \rightarrow 0$) was observed below $T_C$ and magnetism was found to be present in the whole sample volume. The itinerant nature of ferromagnetism in UCoGe is underpinned by the small spontaneous moment of 0.03~$\mu _B$ per U atom. The observation that magnetism in Sr$_{0.5}$Ce$_{0.5}$FBiS$_{2-x}$Se$_x$ for $x \geq 0.5$ is bound to a reduced sample volume also tells us that the small ordered moments measured for $x=0.5$~\cite{Thakur2016} and $x=1.0$ (see SI) are not intrinsic. This naturally explains the `itinerant' behaviour extracted from the magnetization data. In the case of $x=1.0$ the measured moment of $\sim \, 0.2~\mu_B$ can be accounted for by a sizeable ordered Ce moment of $\sim \, 0.7~\mu_B$ in 30~\% of the sample volume. Concurrently, a rough estimate for the magnetic entropy (see SI) associated with the magnetic volume fraction is 1.3$\times$Rln2 (at $T= 10$~K). Thus magnetism keeps its local moment behaviour upon Se doping.

In the case of $x=1.0$ the analysis of the transverse field data shows the superconducting phase occupies the remaining non-magnetic sample volume of $\sim$~70~\% (see Fig.~\ref{figure:fitparTF}a). This value nicely agrees with the superconducting screening fraction deduced from the ac-susceptibility measurements (see Fig.~S4). From the pronounced damping $\sigma _{SC}$ in the superconducting state we can calculate the London penetration depth $\lambda$ with help of the relation $\lambda ^2 \approx 0.0609 \gamma _{\mu} \Phi_0 / \sigma_{SC}$~\cite{Amato1997}. Here $\gamma_{\mu}$ is the muon gyromagnetic ratio ($\gamma_{\mu} /2 \pi = 135.5~$MHz/T) and $\Phi_0$ is the flux quantum. With $\sigma_{SC}=0.70 ~\mu$s$^{-1}$ we calculate $\lambda= 390$~nm for $T \rightarrow 0$. We have also taken transverse field $\mu$SR spectra for the $x=0.5$ compound at a few selected temperatures (data not shown). The data demonstrate superconductivity develops in about 50~\% of the sample volume only, while about 40~\% of the sample is not magnetic and not superconducting even at the lowest temperature $T=0.25$~K.

Notwithstanding our results, the coexistence of superconductivity and ferromagnetism in BiS$_2$-based materials such as Sr$_{0.5}$Ce$_{0.5}$FBiS$_{2}$~\cite{Lin2015} and CeO$_{0.3}$F$_{0.7}$BiS$_2$~\cite{Lee2014} is a remarkable observation and deserves to be studied in detail, notably as regards the possible interplay of local moment magnetism and superconductivity. An important question that has not been answered for these materials yet is the one of electronic phase homogeneity, which calls for $\mu$SR experiments. Our $\mu$SR study on Sr$_{0.5}$Ce$_{0.5}$FBiS$_{2}$ doped with Se irrevocably shows ferromagnetism and superconductivity are phase separated. It provides an excellent example of the power of the $\mu$SR technique in condensed matter physics.

\section*{Methods}
Muon spin relaxation and rotation experiments were carried out with the Multi Purpose Surface Muon Instrument DOLLY installed at the $\pi $E1 beamline at the S$\mu$S facility of the Paul Scherrer Institute (PSI) in Villigen (Switzerland). The $\mu$SR technique makes use of spin-polarized muons implanted in a sample and the ensuing asymmetric decay process into positrons~\cite{Blundell1999}. The positrons are collected in detectors at positions forward and backward with respect to the initial muon spin direction. The muon asymmetry $A(t)$ is determined by calculating $A(t) =(N_B (t) - \alpha N_F (t))/(N_B (t) + \alpha N_F (t) )$, where $N_B (t)$ and $N_F (t)$ are the numbers of positrons detected in the backward and forward detector, respectively, and $\alpha$ is a constant for calibration purposes. The asymmetry function contains detailed information about the spatial distribution of local magnetic fields and their nature, \textit{e.g.} static or fluctuating. By fitting $A(t)$ to model expressions evaluated for different muon relaxation processes~\cite{Yaounc&Dalmas2011} the magnetic properties of the sample can be determined on the microscopic scale. Zero field (ZF) and transverse field (TF) $\mu$SR time spectra were recorded in the longitudinal mode, \textit{i.e.} with the muon spin parallel to the beam direction. In the TF configuration a small magnetic field was applied perpendicular to the beam direction. The samples were attached with General Electric (GE) varnish to the cold finger of a Heliox insert (Oxford Instruments) that allowed for measurements down to $T=0.25$~K. The sample area for the incident muon beam was typically 100~mm$^2$. The $\mu$SR time spectra were analysed with the software package Musrfit~\cite{Suter&Wojek2012} developed at the PSI.
\\
A polycrystalline compound with composition Sr$_{0.5}$Ce$_{0.5}$FBiSSe was obtained by the solid state synthesis procedure as described in Ref.~\citenum{Thakur2016}. Ce$_2$S$_3$, Bi$_2$S$_3$, SrF$_2$, Bi and Se were thoroughly mixed, pelletized and sealed in a quartz tube under vacuum. The tubes were then heated twice at 800~$^{\circ}$C for 24-36 hours with an intermediate grinding. The sample of the Sr$_{0.5}$Ce$_{0.5}$FBiS$_{1.5}$Se$_{0.5}$ compound comes from the same batch as used in Ref.~\citenum{Thakur2016}.  Magnetization, ac-susceptibility, electrical resistivity and specific heat measurements reported in the Supplementary information were carried out in a Physical Property Measurement System equipped with a 9~T superconducting magnet (Quantum Design).

\subsection*{Data availability}
The data needed to evaluate the conclusions in the paper are present in the paper and the Supplementary Information. Additional information may be requested from the authors.

\section*{Acknowledgements}

The authors of Ref.~\citenum{Thakur2016} are gratefully acknowledged for providing a sample of the $x=0.5$ compound. This work was part of a research program funded by FOM (Dutch Foundation for Fundamental Research of Matter). It was also supported by the DFG (Deutsche ForschungsGemeinschaft - GR 4667/1-1 and GRK 1621).

\section*{Author contributions statement}

A.M.N., V.G., R.S., J.-C.O. and A.d.V. conducted the $\mu$SR experiments and analysed the data. M.S. carried out magnetization, ac-susceptibility and electrical resistivity measurements. J.H. carried out specific heat measurements. Y.K.H. synthesized the $x=1.0$ compound. H.-H.K. and A.A. supervised the $\mu$SR experiments. A.M.N. and A.d.V. wrote the manuscript with help of V.G. and R.S. All authors reviewed the manuscript.

\section*{Additional information}

\textbf{Supplementary information} accompanies this paper at http://www.nature.com/srep
\\
\textbf{Competing financial interests} The authors declare no competing financial interests.

\bibliography{Refs_srcefbisse}

\end{document}